\documentclass[11pt]{article}
\pdfoutput=1
\usepackage{geometry}                
\usepackage{graphicx}
\usepackage{amssymb}
\usepackage{amsmath,amssymb}
\usepackage{epsfig}
\usepackage{graphicx}

\usepackage{color}

\newcommand{\be}{\begin{equation}}
\newcommand{\ee}{\end{equation}}
\newcommand{\bea}{\begin{eqnarray}}
\newcommand{\eea}{\end{eqnarray}}

\newcommand{\BM}[1]{{\mbox{\boldmath{$#1$}}}}

\title{
\vspace*{-1cm}
\begin{flushright}
\normalsize{ANL-HEP-PR-11-20\\ }
~\\
~\\
~\\
\end{flushright}
Long Lived Fourth Generation and the Higgs\\
\author{\textbf{Wai-Yee Keung$^{a,b}$ and Pedro  Schwaller$^{a,c}$ } \\~\\
~\\
\normalsize\emph{$^a$ Physics Department, University of Illinois at Chicago, Chicago, IL 60607}\\
\normalsize\emph{$^b$ Department of Physics, Brookhaven National Laboratory, Upton, NY 11973}\\
\normalsize\emph{$^c$ HEP Division, Argonne National Laboratory, 9700 Cass Ave., Argonne, IL 60439}
}}
\begin{document}
\maketitle
\vspace*{2cm}
\begin{abstract}
A chiral fourth generation is a simple and well motivated extension of the standard model, and has important consequences for Higgs phenomenology. Here we consider a scenario where the fourth generation neutrinos are long lived and have both a Dirac and Majorana mass term. Such neutrinos can be as light as 40 GeV and can be the dominant decay mode of the Higgs boson for Higgs masses below the W-boson threshold. We study the effect of the Majorana mass term on the Higgs branching fractions and reevaluate the Tevatron constraints on the Higgs mass. We discuss the prospects for the LHC to detect the semi-invisible Higgs decays into fourth generation neutrino pairs. 
Under the assumption that the lightest fourth generation neutrino is stable, its thermal relic density can be up to 20\% of the observed dark matter density in the universe. This is in agreement with current constraints on the spin dependent neutrino-neutron cross section, but can be probed by the next generation of dark matter direct detection experiments. 
\end{abstract}
\thispagestyle{empty}
\newpage

\section{Introduction}
The fermionic matter content of the standard model of particle physics is organized into three families with chiral couplings to the electroweak SU(2)$\times$U(1) gauge bosons. The minimum of three generations is required by the CKM mechanism for CP violation in the standard model, and has of course been confirmed in the last decades by the discovery of the top quark, the last missing member of the third family of quarks and leptons. On the other hand, there is no immediate upper limit on the number of fermion generations, such that a fourth generation is a well motivated extension of the standard model that should be searched for by experiments. 

Phenomenology of a fourth generation of quarks has been studied exhaustively\footnote{See e.g.~\cite{Frampton:1999xi} for a review, an extensive list of references can also be found in~\cite{Eberhardt:2010bm}.}. Direct searches for additional up and down-type quarks by the Tevatron and very recently at the LHC constrain their masses to be above $335$~GeV and $372$~GeV respectively \cite{Conway:tprime,Aaltonen:2011vr}. 

These searches however depend on the decay modes of the fourth generation quarks~\cite{Flacco:2011ym}. A less model dependent way to constrain fourth generation quarks is through their effects on electroweak precision constraints \cite{He:2001tp,Kribs:2007nz,Novikov:2009kc,Eberhardt:2010bm}
 and on the Higgs production cross section \cite{Kribs:2007nz,Anastasiou:2010bt,Li:2010fu}. The latter has been used recently by the Tevatron to exclude Higgs masses from $131$~GeV-$204$~GeV in the presence of a fourth generation \cite{Aaltonen:2010sv}. 
 
Precise measurements of the $Z$-boson width at LEP excludes the existence of a fourth neutrino flavor with a mass below $M_Z/2$. It follows that the fourth generation neutrino must have a Dirac or a Majorana mass term. In addition the LEP experiments have searched for fourth generation leptons that decay to SM leptons, and also for charged stable leptons, roughly constraining them to be heavier than $100$~GeV. For a recent analysis of these constraints for mixed Dirac Majorana neutrino masses see \cite{Carpenter:2010dt}. Fourth generation Majorana neutrinos are also constrained by limits on lepton flavor violation, assuming a nonzero mixing with the first three generations of fermions \cite{Lenz:2010ha}. 

If the mixing of fourth generation fermions with the standard model is very small, respectively the lightest quark and lepton become long-lived~\cite{Frampton:1997up}. It was recently pointed out that such long-lived quarks not only have distinct phenomenological signatures, but also can assist electroweak baryogenesis scenarios by protecting baryon number from being washed out by sphaleron processes \cite{Murayama:2010xb}. In this case also the LEP constraints on fourth generation neutrinos must be revisited \cite{Carpenter:2010sm}.

A long lived fourth generation neutrino can have important consequences for Higgs phenomenology, since the coupling to the Higgs boson is proportional to the neutrino mass. For $ m_\nu < m_h/2$ the decay into neutrino pairs easily outweighs the Higgs to $b\bar{b}$ width and can be the dominant Higgs decay mode for Higgs masses below $2 M_W$. In section 2 we review the properties of mixed Dirac-Majorana neutrinos, using a two component spinor formalism, and derive the couplings to the Higgs and $Z$-bosons. In section 3 the effects on Higgs production and decay are discussed, paying attention particularly to the effects of the Majorana mass term that splits the two neutrino states. We proceed by discussing the collider phenomenology of Higgs bosons in this scenario in section 4. Finally we calculate the relic density and direct detection constraints on a stable fourth generation Majorana neutrino in section 5, before presenting our conclusions in section 6. In the appendix we translate the results of section 2 to four component spinors. 
\section{Fourth Generation Neutrino Masses and Couplings}
It is instructive to first review the masses and couplings of mixed Dirac-Majorana neutrinos, see also~\cite{Frandsen:2009fs} for a similar analysis. 
We work in the two component spinor formalism, following the conventions of \cite{Dreiner:2008tw}. The Higgs doublet and the fourth generation neutrinos and charged leptons are denoted by
\begin{align}
	\phi = \begin{pmatrix} \phi^+ \\ \phi^0 \end{pmatrix}, 
	\qquad L = \begin{pmatrix}\chi_\nu \\ \chi_\ell \end{pmatrix} \,,
	\qquad \eta_\nu\,,
	\qquad \eta_\ell \,,
\end{align}
with hypercharges ($Y=Q-T_3$) assigned as $+\frac{1}{2}$, $-\frac{1}{2}$, $0$, $+1$ respectively\footnote{Note that all undaggered fields are left-handed, i.e. $\eta_\ell$ is the left-handed positron.} . Gauge invariance further implies that the charged Weyl fields $\chi_\ell$, $\eta_\ell$ can be combined into a Dirac spinor $\psi_{\ell} = (\chi_\ell, \eta_\ell^\dagger)^T$. 

The two types of SU(2) singlets that can be constructed are $\phi^\dagger L$ and $\phi^T \tau L$, with hypercharges $-1$ and $0$, and $\tau = -i \sigma_2$. Therefore the allowed Yukawa couplings are 
\begin{align}
	{\cal L} \supset Y_c \,\phi^\dagger L\, \eta_\ell + Y_n\, \phi^T \tau L \,\eta_\nu\, + \rm{h.c.}
\end{align}
These are the usual charged and neutral lepton Yukawa couplings. Note that we have neglected any mixing with the first three generations of leptons. 
The Higgs field, in unitary gauge, is decomposed as 
\begin{align}
	\phi=\left (0\,,\;  v + \frac{1}{\sqrt{2}} h\right )^T\,,
\end{align}
with $v = \sqrt{2}m_W/g \approx 174$~GeV. 

Here we are mainly interested in the interactions of the physical Higgs boson with the neutrinos. Dropping the subscripts on $\chi_\nu$ and $\eta_\nu$, the relevant part of the Lagrangian becomes
\begin{align}
\left( m_{4D} + \frac{m_{4D}}{\sqrt{2}v} h \right) \chi \eta +
\left( m_{4D} + \frac{m_{4D}}{\sqrt{2}v} h \right) \chi^\dagger \eta^\dagger + \frac{1}{2} M\left( \eta \eta + \eta^\dagger \eta^\dagger \right),
\end{align}
where we have introduced the Majorana mass term, and defined $m_{4D} = Y_n v$. The resulting mass matrix can be diagonalized using the Takagi decomposition \cite{Dreiner:2008tw}. Note that in general the masses are not the eigenvalues of the mass matrix ${\cal M}$, but the positive square roots of the eigenvalues of ${\cal M}^\dagger {\cal M}$. The masses and corresponding eigenstates are given by
\begin{align}
	M_{1,2} & = \sqrt{\frac{M^2}{4} + m_{4D}^2} \pm \frac{M}{2}\,,   \\
	N_1 & = i\left( c_\theta \chi - s_\theta \eta\right) \,,\label{eqn:N1}\\
	N_2 & = s_\theta \chi + c_\theta \eta \,,\label{eqn:N2}
\end{align}
with 
\begin{align}
	\tan\theta = \frac{m_{4D}}{M_2} = \frac{M_1}{m_{4D}} = \sqrt{\frac{M_1}{M_2}}\,.
\end{align}
The physically relevant quantities are the two Majorana masses $M_{1,2}$, which we take as the two independent free parameters of the model. The coupling to the Higgs boson, which is proportional to $m_{4D}$, can be rewritten in terms of $M_{1,2}$ as follows:
\begin{align}
	m_{4D} & = \frac{1}{2}\sqrt{(M_1+M_2)^2 - (M_2-M_1)^2} = \sqrt{M_1 M_2} = M_1 \sqrt{\frac{M_2}{M_1}}\,.
\end{align}
Note that for fixed $M_1$, which is bounded by LEP to be larger than about $40$~GeV, it increases with the square root of the ratio of the masses. To remain in the perturbative regime, we require $m_{4D} \lesssim 400$~GeV, which implies that $M_2 \lesssim 4$~TeV. 

The couplings to the Higgs boson are now obtained using equations (\ref{eqn:N1}) and (\ref{eqn:N2}):
\begin{align}
	{\cal L} = \frac{m_{4D}}{\sqrt{2}v} h\left ( \chi \eta + {\rm h.c.} \right) = \frac{1}{2}\frac{m_{4D}}{\sqrt{2}v} h \left (2 c_\theta s_\theta N_1^2 +  2 c_\theta s_\theta N_2^2 +2  i(s_\theta^2 - c_\theta^2) N_1 N_2 + {\rm h.c.}   \right). \label{eqn:higgscoupling}
\end{align}
We have extracted factor of $1/2$ to obtain the canonical normalization for the Majorana-Higgs couplings.
In terms of $M_{1,2}$, the coupling of the lightest neutrino to the Higgs is given by
\begin{align}
	\frac{m_{4D}}{\sqrt{2}v} 2 c_\theta s_\theta = \frac{M_1}{\sqrt{2} v} \frac{2 M_2}{M_1 + M_2} \,.
\end{align}
In the limit $M_2 \to M_1$, the sum of the partial widths should agree with the width of the Higgs decaying into a single Dirac neutrino. The sum of the widths is proportional to
\begin{align}
	\frac{1}{2} \left( 4 c_\theta^2 s_\theta^2 + 4 c_\theta^2 s_\theta^2 \right) + (c_\theta^2 - s_\theta^2)^2 =  (c_\theta^2 + s_\theta^2)^2 = 1 \,, 
\end{align} 
so it is independent of the mixing angle, in the limit where the phase space factors can be neglected, i.e. for $m_h \gg 2 M_2$. The factor $1/2$ in front of the first term is from the phase space for identical particles. The $1/2$ in the coupling (\ref{eqn:higgscoupling}) is compensated for the $N_1N_1$ and $N_2N_2$ channels since each field can be contracted with each final state. For the $N_1 N_2$ channel the factor $1/2$ cancels the factor of two in the coupling, such that the relative factors emerge as shown above.  

For completeness, we also give the couplings to the Z boson. In two component notation, they are obtained from
\begin{align}
	i Y_L \frac{g}{c_w}(\chi^\dagger \bar\sigma^\mu \chi) Z_\mu & = \frac{i g}{2 c_w} Z_\mu \left( c_\theta^2 N_1^\dagger \bar\sigma^\mu N_1 + s_\theta^2 N_2^\dagger \bar\sigma^\mu N_2 + i c_\theta s_\theta( N_1^\dagger \bar \sigma^\mu N_2 - N_2^\dagger \bar\sigma^\mu N_1 )\right)\notag  \\
	& = \frac{i g}{2 c_w} Z_\mu \left( c_\theta^2 N_1^\dagger \bar\sigma^\mu N_1 + s_\theta^2 N_2^\dagger \bar\sigma^\mu N_2 + i c_\theta s_\theta( N_1^\dagger \bar \sigma^\mu N_2 + N_1 \sigma^\mu N_2^\dagger )\right) \,.
\end{align}
In four component notation, this translates into axial couplings of the $Z$ boson to $N_1N_1$ and $N_2N_2$ and a vector coupling to $N_1 N_2$, as discussed in more detail in the appendix. 

Before moving to the next section, let us briefly comment on the phenomenology of the fourth generation charged leptons in this model. Direct searches at LEP constrain the mass of the charged lepton to be larger than 100.8~GeV~\cite{Nakamura:2010zzi}, while electroweak precision tests constrain the mass difference between the charged and neutral lepton, $|m_{\ell_4} - m_{\nu_4}|\lesssim 140~{\rm GeV}$~\cite{Eberhardt:2010bm}, for a Dirac $\nu_4$. 

Pair production of $\ell_4 \bar\ell_4$ pairs with subsequent decays $\ell_4 \to W N_{1,2}$ lead to signatures with leptons, jets and missing energy. In the presence of a nonzero Majorana mass term, same sign lepton pairs can be produced in the process
\begin{align}
	pp \to \ell_4^\pm \ell_4^\pm + X \to \ell^\pm \ell^\pm + E\!\!\!/ + X\,,
\end{align}
where $X$ denotes the proton remnants from the $W$ fusion process. In addition charged-neutral pairs $\ell_4 N_{1,2}$ can be produced and lead to signals with one charged lepton and missing energy. Some of these processes were considered e.g. in~\cite{Carpenter:2010dt,Frandsen:2009fs}. For a recent study of same sign leptons see also~\cite{Atre:2009rg}, however in our case the stable $N_1$ and the neutrinos from $W$ decays lead to much larger missing energy. 
\section{Higgs Production and Decay}
Higgs production through gluon fusion is enhanced  in the presence of a heavy fourth generation. For $m_h \leq 200$~GeV, the enhancement is well approximated by multiplying the standard model production rate by a factor $n_h^2 = 9$ \cite{Kribs:2007nz,Anastasiou:2010bt,Li:2010fu}, and is largely independent of the structure of the lepton sector. If Higgs decays to the fourth generation are not allowed kinematically, this greatly strengthens the bounds on the Higgs mass from the Tevatron experiments, excluding the $131-204$~GeV region \cite{Aaltonen:2010sv}.

\begin{figure}
\begin{center}
\includegraphics{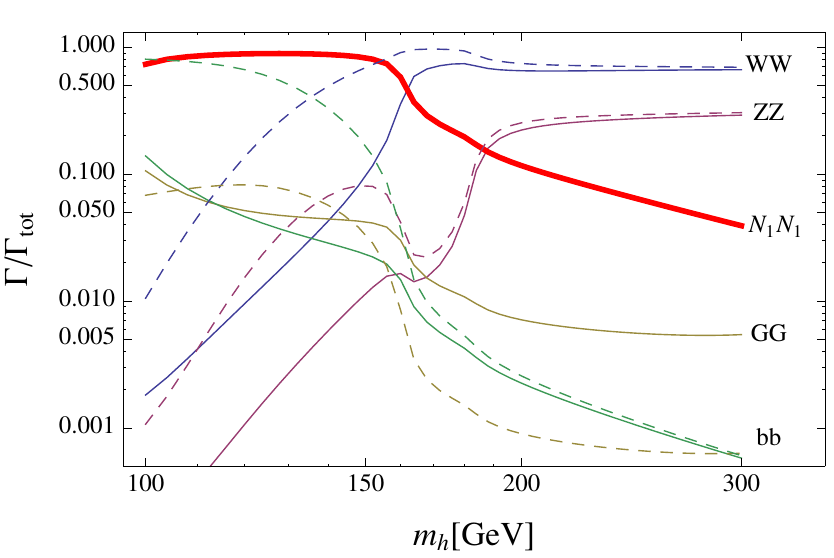}
\caption{Higgs branching fractions as function of the Higgs mass. Solid lines correspond to a fourth generation Dirac neutrino with mass $M_1= 45$~GeV, dashed lines are the corresponding standard model branching fractions. Above the $h\to WW$ threshold the branching fractions converge to their corresponding standard model values, except for the gluon channel that remains enhanced. }
\label{fig:hdec1}
\end{center}
\end{figure}

The light Higgs mass range can be resurrected in the presence of a long lived fourth generation due to the invisible decays $h \to \bar{\nu}_4 \nu_4$. The tree level decay width into a fourth generation Dirac neutrino with mass $M_1$ is given by
\begin{align}
	\Gamma_{\nu\nu}=\Gamma(h \to \bar{\nu}_4 \nu_4) & = \frac{1}{16 \pi} \frac{M_1^2}{v^2} m_h\left( 1- \frac{4 M_1^2}{m_h^2} \right)^{3/2}\,.
\end{align}
For a light Higgs this competes with the decay into b quarks, and the ratio of the width is approximately given by
\begin{align}
	\frac{\Gamma_{\nu 4}}{\Gamma_b} \approx \frac{M_1^2}{3 m_b^2}\,.
\end{align}
The decay into fourth generation neutrinos therefore dominates the total width of the Higgs. 
In Figure (\ref{fig:hdec1}) we show the branching fractions of the Higgs boson in the presence of a purely Dirac neutrino $\nu_4$ with a mass of 45~GeV. We have also included the enhancement of the $h\to gg$ channel due to the fourth generation quarks. Our findings for the case of a Dirac neutrino agree with the results of previous studies \cite{Belotsky:2002ym,Rozanov:2010xi}. The SM branching fractions of the Higgs were obtained using HDECAY~\cite{Djouadi:1997yw}.

The situation becomes more complicated in the presence of  a nonzero Majorana mass. In this case, we find that the partial widths are given by
\begin{align}
	\Gamma(h \to N_1 N_1) & = \frac{1}{2}\frac{1}{16 \pi} \frac{M_1^2}{v^2} \frac{(2M_2)^2}{(M_1+M_2)^2}\,m_h\left( 1- \frac{4 M_1^2}{m_h^2} \right)^{3/2}, \\
	\Gamma(h \to N_2 N_2) & = \frac{1}{2}\frac{1}{16 \pi}\frac{M_1^2}{v^2} \frac{(2M_2)^2}{(M_1+M_2)^2} \, m_h\left( 1- \frac{4 M_2^2}{m_h^2} \right)^{3/2}, \\
	\Gamma (h \to N_1 N_2) & = \frac{1}{16 \pi} \frac{M_1 M_2}{v^2} \frac{(M_2 - M_1)^2}{(M_2 + M_1)^2} \, m_h \left( 1- \frac{(M_2-M_1)^2}{m_h^2}\right) \sqrt{\lambda\left(\frac{M_1}{m_h},\frac{M_2}{m_h}\right)}\,.
\end{align}
The definition for the triangle function $\lambda$ can be found in the appendix. 
As discussed above, in the Dirac limit, where $M_2 \to M_1$, the width into $N_1 N_2$ goes to zero, and $\Gamma_{11} + \Gamma_{22} = \Gamma_{\nu \nu}$. On the other hand, in the limit where $M_2 \gg M_1, m_h$, only the decay into $N_1 N_1$ is possible. It is interesting to note that in this case the factor $1/2$ in the width is overcompensated by the factor $(2M_2)^2/(M_1+M_2)^2 \to 4$, and we have that
\begin{align}
	\Gamma (h \to N_1 N_1) \big\vert_{M_2\gg M_1} & = 2 \Gamma(h \to \bar{\nu}\nu). 
\end{align}
This can also be understood in terms of an effective theory. Integrating out the right-handed neutrino, the Majorana mass for the left-handed neutrino comes from a term 
\begin{align}
\frac{1}{2}\frac{Y_n^2}{M_2}(\phi L)(\phi L) = \frac{1}{2}\frac{M_1}{v^2} (\phi L)(\phi L)  \to \frac{1}{2} M_1 N_1 N_1 + \frac{1}{2} \frac{\sqrt{2} M_1}{v} h N_1 N_1 \,,
\end{align}
i.e. the coupling to the Higgs is enhanced by a factor of two compared to the case of a Dirac mass term.

We can therefore conclude that the presence of a Majorana mass term in the fourth generation neutrino sector can increase the invisible width of the Higgs boson by up to a factor of two. It is decreased compared to the Dirac case in a narrow region of parameter space, where $M_2$ is just large enough to forbid Higgs decays into $N_2N_2$. The suppression is 50\% at most, i.e. $\Gamma_{11}> \Gamma_{\nu\nu}/2$. 

\begin{figure}
\begin{center}
\includegraphics[width=.49\textwidth]{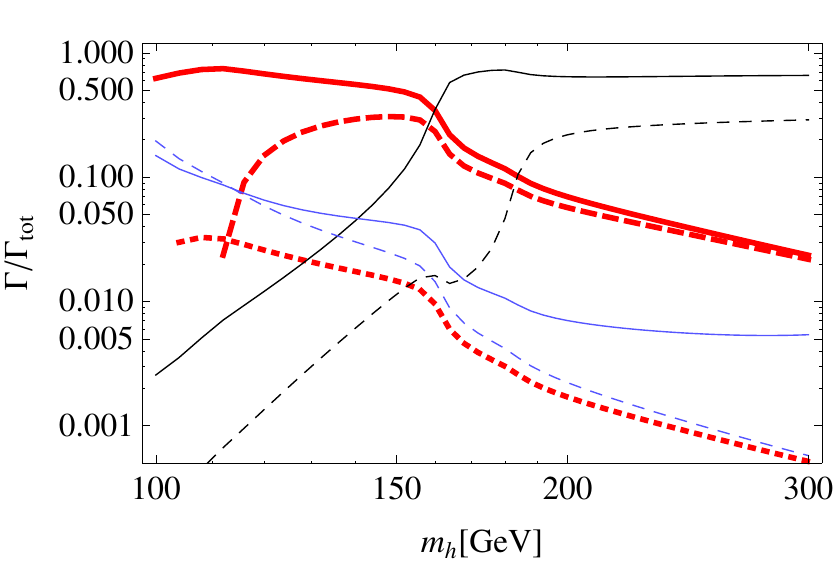}
\includegraphics[width=.49\textwidth]{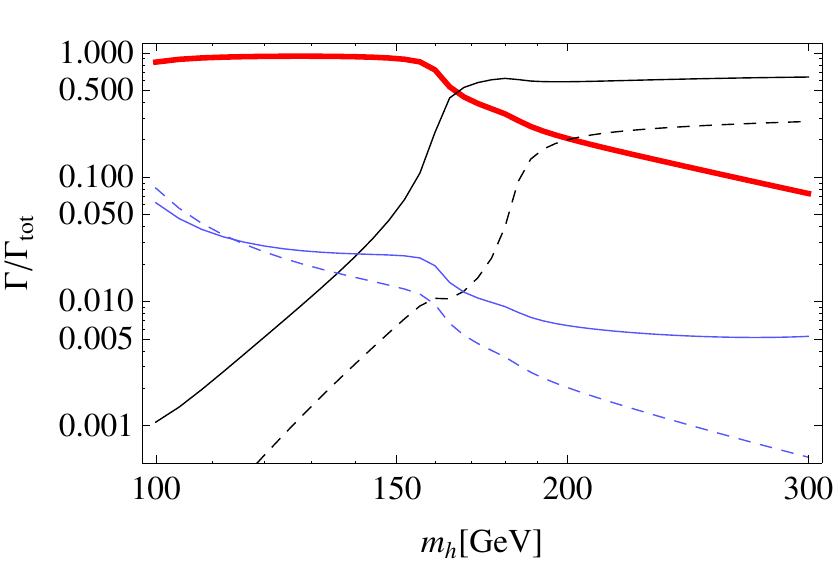}
\caption{Higgs branching fractions in the presence of stable fourth generation Majorana neutrinos. In the left plot we assume a small mass splitting, $M_1 = 45$~GeV and $M_2=55$~GeV. The right plot shows the limit where $N_2$ decouples from the theory. \newline
Thick red lines are the  Higgs branching fractions into $N_1 N_1$ (solid), $N_2 N_2$ (dashed) and $N_1 N_2$ (dotted). The black lines are the $WW$ (solid) and $ZZ$ (dashed) channels, while the blue (light grey) lines are the $GG$ (solid) and $bb$ (dashed) channels. Decays into charm, strange and tau pairs are not shown, other decays have branchings smaller than $5\times 10^{-3}$. }
\label{fig:hdec2}
\end{center}
\end{figure}

To illustrate the effects of a nonzero Majorana mass, in figure \ref{fig:hdec2} we show the branching fractions of the Higgs boson for two characteristic cases. We always take $M_1 = 45$~GeV to simplify the comparison with the Dirac case. For the left plot we take $M_2 = M_1 + 10$~GeV such that all decay channels are open. The right plot is valid in the limit where $N_2$ decouples, i.e. $M_2 \gg M_1$. 

Comparing with the Dirac case, one can see that the suppression of the  standard model Higgs decays is somewhat weaker in the presence of a small Majorana mass, while it is stronger for the case of a large Majorana mass term. The $N_2N_2$ channel suffers from phase space suppression compared to the $N_1N_1$ channel, but will be comparable whenever $m_h \gg 2 M_2$. The $N_1N_2$ channel is always suppressed, either by the $(M_2 - M_1)^2$ factor in the width, or by phase space when $M_2 \gg M_1$. For moderate mass splittings $M_2 - M_1 \sim 20$~GeV the branching fraction can reach up to 8\%, for Higgs masses below the $WW$ threshold. 

\section{Higgs Phenomenology with mixed $N_1$, $N_2$}
The limit on the neutrino mass of $M_1 > 45$~GeV only applies for stable Dirac neutrinos. For a pure Majorana $N_1$ instead one obtains $M_1 > 39.6$~GeV \cite{Nakamura:2010zzi}. 
In the mixed case the constraints on $M_1$ can be further reduced, since the coupling to the $Z$ boson can be suppressed by the mixing angles. In addition one has to take into account LEP constraints on $N_2 \to N_1 + {\rm X}$ decays. This was recently analyzed in \cite{Carpenter:2010sm}. The allowed regions for $M_1$, $M_2$ are roughly given by:
\begin{align}
	&\rm{I} &&M_1> 30~\rm{GeV} && M_2 - M_1 < 20~\rm{GeV}\,,  \\
	&\rm{II} && M_1 \gtrsim 40~\rm{GeV} && M_2 + M_1 \gtrsim 170~\rm{GeV}\,.
\end{align}
For previous studies of Majorana neutrino production through Higgs boson exchange, see also~\cite{Datta:1991mf,Datta:1993nm} and~\cite{Frandsen:2009fs}.
\subsection{Higgs Mass Bounds}
In the presence of a fourth generation, the production of Higgs bosons in gluon fusion is enhanced by roughly a factor of 9 \cite{Kribs:2007nz}. On the other hand the associated production of a Higgs with a $Z$ or $W$ boson is not significantly changed.

\begin{figure}
\begin{center}
\includegraphics{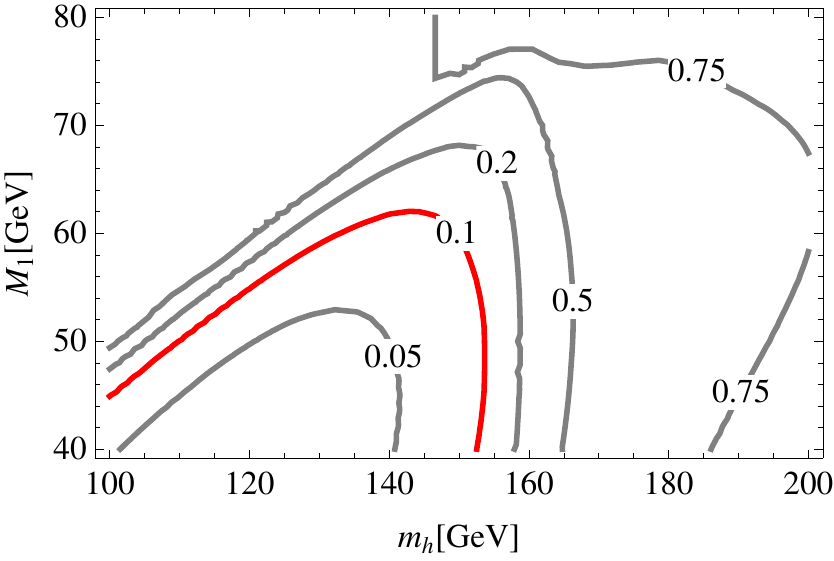}
\caption{Suppression of the branching fraction $\rm{Br}(h \to WW^*)$ relative to the SM branching fraction, as a function of $m_h$ and $M_1$, for $M_2 = 4000$~GeV. Shown are the contours for the ratio $\rm{Br}_{\rm 4G}(h \to WW^*) / \rm{Br}_{\rm SM}(h \to WW^*)$.
Below the red contour, the suppression is stronger than the enhancement of Higgs production in gluon fusion. The suppression in the region where the decay $h\to N_1 N_1$ is not allowed is due to the fourth generation enhancement of the Higgs decay to gluon pairs. 
}
\label{fig:brww}
\end{center}
\end{figure}

The LEP experiments pose stringent bounds on the invisibly decaying Higgs boson, such that $m_h > 114$~GeV continues to hold in any scenario considered here. In the low mass region, $m_h < 130$~GeV, the strongest constraints come from $ZH$ and $WH$ channels. These limits are not enhanced by the increased gluon fusion rate, such that the low mass region remains viable. 

Above $m_h = 130$~GeV, the $p\bar{p}\to h\to WW^*$ mode is strongly enhanced and, in the absence of new Higgs decay channels, excludes a Higgs in the mass range $131~{\rm GeV} < m_h < 204~{\rm GeV}$ \cite{Aaltonen:2010sv}. The suppression of the $WW$ width from Higgs decays into fourth generation neutrinos is crucial to resurrect Higgs masses between $131$~GeV and $2 M_W$. The relevant condition is
\begin{align}
	9  \times \frac{{\rm Br}_{4G}(h \to WW^*)}{{\rm Br}_{\rm SM}(h \to WW^*)} \leq \sigma_{h,\rm excluded}
\end{align}
where $\sigma_{h,excluded}$ is the Higgs production cross section excluded by the combined Tevatron limit, normalized to the standard model production cross section. The suppression of the $h\to WW$ branching fraction for different values of $M_1$ is shown in Fig.~\ref{fig:brww} for the case when $M_2$ decouples. 
For $M_1 = 50$~GeV, we obtain an upper bound on the light Higgs mass of
\begin{align}
	m_h < 156~{\rm GeV} \,,
\end{align}
which is valid for $M_2 \gg M_1$, i.e. for a large Majorana mass term. This result is quite stable under small variations of $M_1$. Lowering the mass decreases the coupling to the Higgs but enlarges the available phase space, while increasing the mass has the opposite effect. This can also be seen in Fig.~\ref{fig:brww}, where the contours of equal suppression are approximately straight vertical lines, away from the threshold for $h\to N_1 N_1$. In the regime where $M_2 - M_1 \ll M_2$, i.e. for a small Majorana mass,  the $h\to WW^*$ width is slightly less suppressed, such that Higgs masses down to 152~GeV become excluded. 

In the large Higgs mass region, the branching into on-shell $WW$ pairs is only slightly reduced by the neutrino decay modes, such that the lower bound for the high mass region is not significantly changed from $m_h > 204$~GeV obtained in \cite{Aaltonen:2010sv}.

In region I the decays $h \to N_1 N_2$ and $h\to N_2N_2$ are possible, and offer a possibility to discover the otherwise invisible decays of the Higgs bosons through the decays of $N_2$ to $N_1$, through an off-shell $Z$ or Higgs boson. 
\subsection{$N_2$ Decays}
In addition to the decays $N_2 \to N_1 Z^{(*)}$ that have been studied in the literature, one must also take into account decays of $N_2$ into on- or off-shell Higgs bosons. The relevant couplings to the $Z$ and Higgs are 
\begin{align}
	\frac{g}{2 c_w} c_\theta s_\theta &= \frac{M_Z}{\sqrt{2} v} \frac{2M_2}{M_1 + M_2} \\
	\frac{m_{4D}}{\sqrt{2}v} 2 (c_\theta^2 - s_\theta^2) & = \sqrt{2} \frac{M_1}{v} \frac{M_2-M_1}{M_2+M_1}
\end{align}
When the mass splitting is small, the coupling to the Higgs is suppressed. In addition, since both decays are into virtual particles here, the small width of a light Higgs boson makes this channel even more suppressed. When $M_2$ is large enough such that both decays can happen on-shell, we will roughly have that $\Gamma_h / \Gamma_Z \sim M_1^2/M_Z^2$. In this regime the Higgs will mostly decay invisibly. 
We can therefore conclude that, as far as Higgs physics is concerned, it suffices to consider the decays $N_2 \to N_1 Z$. 

For direct $N_2$ searches at the LHC, other parameter regions might become important. In particular for $M_2 -M_1 > m_h$ and either $m_h \sim 155$ or $m_h > 200$, we will have a situation where the Higgs decay to $W^+W^-$ is significant. This can lead to the following interesting signals:
\begin{align}
	pp \to &N_1 N_2 \to W^+ W^- + {E\!\!\!/} \\
	pp \to & N_2 N_2 \to W^+ W^- Z + {E\!\!\!/}
\end{align}
These channels might offer improved sensitivity for LHC searches in the high mass regime. With production cross sections in the $(10-100)$~fb range at the $\sqrt{s}=7$~TeV LHC it will still be a very challenging search. 
\subsection{Higgs Phenomenology with Small $M_2$}
We first consider the light Higgs case. 
Since $M_2 - M_1 < 20$~GeV,  both $h\to N_1N_2$ and $h\to N_2 N_2$ are kinematically accessible. $N_2$ decays to $N_1$ via an off-shell $Z$ boson, while decays mediated by an off-shell Higgs are suppressed. 
The most promising channel therefore is
\begin{align}
	N_2 \to \ell^+ \ell^- + {E\!\!\!/}\,.
\end{align} 
One can look for the Higgs boson in the di-lepton and four lepton channel with missing energy. For Higgs bosons produced in the gluon fusion mode, the leptons will have very small transverse momenta, and fail most experimental cuts. 
It is more promising to consider associated $Zh$ and $Wh$ production. Not only will the $Z$ and $W$ bosons provide a trigger, we can also expect a moderate boost for the Higgs, such that the final state leptons pass some mild $p_T$ cuts. 

The production cross section for $Zh$ and $Wh$ at the Tevatron, for a light Higgs, are ${\cal O}(100~{\rm fb})$. Channels with the Higgs going to four leptons are suppressed too much by the $Z$ boson branching fractions. The most promising channel is 
\begin{align}
	Wh \to \ell \nu_\ell + \ell^+ \ell^- {E\!\!\!/}\,,
\end{align}
with a cross section of ${\cal O}(1~{\rm fb})$ at the Tevatron. The most recent measurements from the D0 and CDF experiments constrain the cross section for $pp\to 3\ell + {E\!\!\!/}$ to be smaller than $0.1$~pb, still two orders of magnitude above the expected signal. 

At the LHC with $\sqrt{s}=7$~TeV the $WH$ and $ZH$ production cross sections can reach up to $0.5$~pb. Backgrounds for this channel come from $WZ$ and $W\gamma^*$ production as well as from $t\bar{t}$ production, where additional leptons are produced in decays of heavy flavors. A recent CMS study~\cite{CMSmultilepton} shows that, with a moderate cut on missing transverse energy, ${E\!\!\!/}_T>50$~GeV, and requiring that the invariant mass of same flavor lepton pairs is away from the $Z$~pole, the backgrounds can be reduced to the 30~fb level, comparable to the expected signal rates.

 Finally the conventional Higgs search channels remain viable, although modified by a factor $9\times \Gamma_{4G}/\Gamma_{\rm SM}$. In the case that the Higgs is found in one of those channels, the measured production cross section will be different from the standard model prediction. It is then crucial to look for the semi-invisible Higgs decays to understand the origin of this deviation from standard model predictions. 
\subsection{Higgs Phenomenology with Large $M_2$}
Here we assume that we are in case (II), but without decoupling $N_2$. In the small $m_h$ region, the Higgs decays only to $N_1N_1$, which is invisible. Finding the Higgs in this regime will be difficult, since the conventional search channels are significantly suppressed, in particular those based on associated Higgs production. The invisibly decaying Higgs may however be observed in the weak boson fusion channel at the LHC~\cite{Eboli:2000ze}. 

For the region $200~{\rm GeV} < m_h < 300~{\rm GeV}$, the Higgs decays into pairs of electroweak gauge bosons dominate. The $N_1 N_2$ and $N_2 N_2$ channels suffer from phase space suppression in this regime, such that their branching fractions are only at the $1\%$ level. 

The $N_2$ now decays into an on-shell $Z$ boson. The relevant leptonic channels are
\begin{align}
	pp \to &h \to \ell^+ \ell^- + {E\!\!\!/} \,,\\
	pp \to &h \to \ell^+ \ell^- \ell^+ \ell^- + {E\!\!\!/}\,.
\end{align}
The di-lepton channel receives contributions both from $N_1N_2$ and $N_2N_2$ decays, when one of the $Z$ bosons decays invisibly. The dominant SM backgrounds for this channel are $ZZ$ and $WW$ pair production. In addition, there is also an irreducible background from electroweak $N_1N_2$ and $N_2 N_2$ production, mediated by a $Z$ boson. The signal rate for this channel at the $7$~TeV LHC is about
\begin{align}
	{\cal O}(10^{-3} ) \times \sigma(gg \to h)_{\rm 4G} \sim {\cal O}(10-100)\,{\rm fb} \,,
\end{align}
while it is largely invisible at the Tevatron. At the LHC the background from $W^+W^-$ production can be reduced by requiring the invariant mass of the lepton pair to be close to $m_Z$. However the background from $ZZ \to \ell^+\ell^- + {E\!\!\!/}$ is irreducible and, with about $170$~fb~\cite{Campbell:2011bn}, a few times larger than the potential signal. This background can be measured accurately at the LHC using $ZZ\to 4\ell$ decays. Direct $N_1 N_2$ production contributes ${\cal O}(20~{\rm fb})$ to the $\ell^+\ell^-+{E\!\!\!/}$ background. With $5$~fb$^{-1}$ at the 7~TeV LHC a significant excess in this channel might be observable, depending on the precise values of $m_h$ and $M_{1,2}$. 
\section{Fourth Generation Neutrino Dark Matter }
For the above analysis to be valid, it is sufficient if the $N_1$ lifetime is long enough to escape the LEP detectors before they decay. However it is also conceivable that $N_1$ is stable on cosmological timescales, and thus contributes to the dark matter of the universe. 

It is well known that stable fourth generation Dirac neutrinos make very bad dark matter candidates. The vectorlike coupling to the Z-boson results in an unsuppressed annihilation cross section into fermion pairs for low momenta, and in a large elastic scattering cross-section with nucleons. The latter strongly constrains the relic density of Dirac neutrinos to be several orders of magnitude smaller than the observed dark matter density. 

Some of these problems can be avoided when the Dirac pair is split using a Majorana mass term. In that case the axial coupling of $N_1$ to the $Z$ boson helps avoiding most constraints from direct detection, provided that the mass splitting is larger than the typical ${\cal O}$(100~keV) momentum transfer in dark matter nucleon scatterings. 
See e.g.~\cite{TuckerSmith:2001hy} for a discussion in the context of inelastic dark matter.
\subsection{Majorana Neutrino Annihilation to Fermions}
The annihilation rate of a Majorana fermion into standard model fermion pairs through $Z$-boson exchange is well known from the case of neutralino annihilation in the MSSM. Here we briefly review the most important results. The coupling of $N_1$ to the $Z$-boson in four component language is proportional to $\gamma_\mu \gamma_5$. We consider the annihilation into a pair of SM fermions with mass $m_f$ and with a coupling to the $Z$ boson of the form
\begin{align}
	\frac{e}{2 s_w c_w} \left( T_3 \gamma_\mu (1-\gamma_5) - 2 Q_f s_w^2 \gamma_\mu \right)\,,
\end{align}
where $Q_f$ is the fermion electric charge and $T_3$ is the isospin eigenvalue. We work in the center of mass frame, where the $N_1$ three-momentum can be replaced by $|\mathbf{p}| = \sqrt{s/4-M_1^2}$. 
The spin-averaged squared matrix element, in unitary gauge, and integrated over solid angles, is then given by:
\begin{align}
	\int \frac{d\Omega}{4\pi} & \frac{1}{4}\sum|{\cal M}|^2 = \frac{2 c_t^4 e^4}{6 c_w^4 s_w^4 M_Z^4}\,\frac{1}{(s-M_Z^2)^2+M_Z^2 \Gamma_Z^2}\times \Big[\,
	 \frac{3}{4} M_1^2 m_f^2 (s - M_Z^2)^2 +  \\
	&+ s^2 \left(1-\frac{4M_1^2 }{s}\right)\left(-\frac{m_f^2}{s} +1 + 8Q_f^2 \left( 1+\frac{2 m_f^2}{s} \right)s_w^4 \mp
    4 Q_f \left(1+\frac{2 m_f^2}{s}  \right) s_w^2 \right) \Big],
  \notag
\end{align}
where we used that $T_3 = \pm 1/2$.
This result displays the important helicity suppression for this process: In the $|\mathbf{p}|\to0$ limit the amplitude is suppressed by the small fermions mass $m_f$ coming from the first term, while the second term vanishes as $|\mathbf{p}|^2$. Further note that the momentum independent part is not resonantly enhanced at $s = M_Z^2$ since the pole of the $Z$ propagator is cancelled by the numerator. 

From this the annihilation cross section is obtained using
\begin{align}
	\langle v\sigma \rangle & = \frac{1}{8 \pi s} \left( 1 - \frac{4 m_f^2}{s}\right)^{1/2} \int \frac{d\Omega}{4 \pi}\frac{1}{4} |{\cal M}|^2\,.
\end{align}
Freeze-out typically occurs at temperatures $T \sim M_1/20$, such that the majority of $N_1$ are non-relativistic.  In this limit the center of mass energy can be approximated by $s = 4 M_1^2 + v^2 M_1^2$, where $v$ is the relative velocity between the annihilating $N_1$. Away from the $Z$ pole region we can further set $\Gamma_Z \to 0$. Decomposing the annihilation cross section as $\langle v \sigma \rangle = a + b v^2$, and dropping terms of order $m_f^2 v^2$  and $v^4$, we obtain the coefficients
\begin{align}
	a &= \frac{1}{128 \pi} \frac{c_\theta^4 g^4}{c_w^4} \frac{1}{M_Z^4} \sum N_f m_f^2\,, \\
	b & = \frac{c_\theta^4 g^4}{192 \pi c_w^4}\frac{M_1^2}{(4 M_1^2 - M_Z^2)^2} \sum_f \left( 1 - {\rm sign}(T_{3f}) 4 Q_f s_w^2 + 8 s_w^4 Q_f^2 \right) \notag \\
	& = \frac{1}{192\pi} \frac{c_\theta^4 g^4}{c_w^4}  \frac{M_1^2}{(4 M_1^2 -M_Z^2)^2} \left( 21 - 40 s_w^2 + \frac{160}{3} s_w^4 \right)\,,
\end{align}
valid for $M_1 \gg m_f$, and with $N_f$ denoting the internal degrees of freedom of the fermion $f$. The result for $b$ includes the sum over three families of quarks and leptons, excluding the top quark. The $a$ term is suppressed by $m_b^2/M_Z^2$ relative to the $b$ term, so that even at temperatures around the freeze-out the $b$ term will dominate the annihilation. 

A good first estimate for the relic density can be obtained using \cite{Servant:2002aq}
\begin{align}
	\Omega_{N_1} h^2 \approx \frac{1.04\times 10^9~{\rm GeV}^{-1}}{M_{\rm pl}} \frac{x_F}{\sqrt{g_\star}} \frac{1}{a + 3b/x_F}\,,
\end{align}
where the freeze-out temperature $x_F = M_1/T_F$ is determined numerically from
\begin{align}
	x_F = \log \left[ c(c+2)\sqrt{\frac{45}{8}} \frac{g_d}{2 \pi^3} \frac{M_1 M_{\rm pl} (a + 6b/x_F)}{\sqrt{g_\star}\sqrt{x_F}} \right]\,.
\end{align}
For our case we have $g_d=2$, $g_\star = 92$ and we use $M_{\rm pl} = 1.22\times 10^{19}$~GeV. The resulting relic density reaches at most ${\cal O}(10\%)$ of the observed dark matter abundance in the universe. For $M_1 \sim M_Z/2$ and $M_1 \sim m_h/2$ the relic density is strongly suppressed by resonant annihilation into $Z$ bosons or Higgs bosons respectively, while for $M_1 > M_W$ annihilation into gauge boson pairs becomes dominant and further reduces the relic density. 

\begin{figure}
\begin{center}
\includegraphics{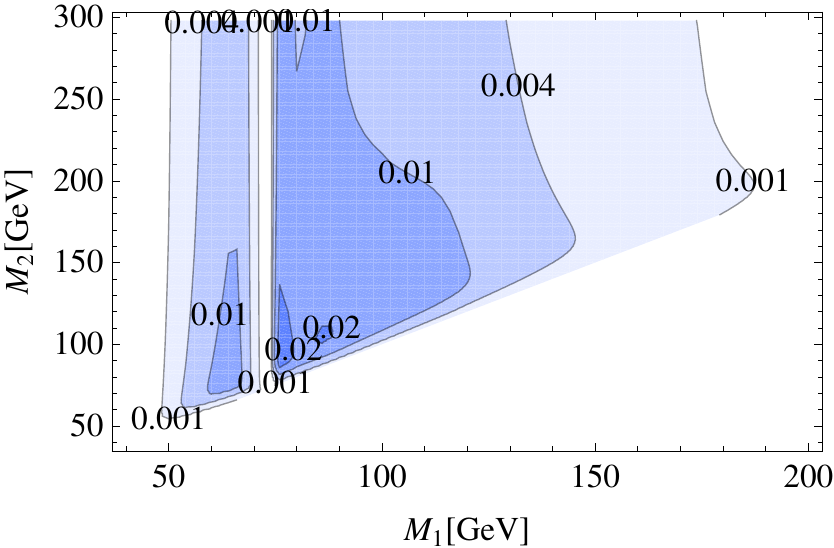}
\caption{$N_1$ relic density $\Omega_{N_1} h^2$ in the $M_1$-$M_2$ plane, for $m_h = 150$~GeV and $M_\ell = M_1 + 200$~GeV. For comparison, the observed dark matter density is $\Omega_{\rm DM} h^2 \approx 0.11$. }
\label{fig:rd}
\end{center}
\end{figure}

In fig. \ref{fig:rd} we show the $N_1$ relic density in the $M_1-M_2$ plane. To capture all possible annihilation channels and to include coannihilation effects, the model was implemented in micrOMEGAs 2.4 \cite{Belanger:2010gh}. Clearly visible are the regions of resonant annihilation near $M_1 = 45$~GeV and $M_1 = 75$~GeV (the Higgs mass was set to $m_h = 150$~GeV). Furthermore the relic density is reduced close to the $M_2 = M_1$ line due to coannihilation, and for increasing $M_1$ where annihilation into gauge boson and top quark pairs becomes possible. 
\subsection{Direct Detection of Majorana Neutrino Dark Matter}
The stringent direct detection constraints on Dirac neutrino dark matter are evaded provided that $M_2 - M_1> 100$~keV \cite{TuckerSmith:2001hy}. Furthermore if we assume that the $N_1$ abundance is purely thermal, then the direct detection rate has to be multiplied by the fraction of dark matter that is of $N_1$ type. In this case we find that the latest CDMS and Xenon 100 results do not impose a constraint on the parameter space. 

In addition one can consider the case where all dark matter consists of $N_1$. This can be achieved by reducing the coupling of $N_1$ to the Z-boson, as suggested in \cite{Belanger:2007dx}. Here we do not consider this possibility since this also affects the direct detection rates. Another way to increase the relic density of $N_1$ is to invoke a nonthermal production mechanism, e.g. using a heavy scalar that couples to the standard model only through a very small coupling to $N_1$, and decays to $N_1$ pairs after freeze-out.  It requires a somewhat accurate adjustment of masses and couplings to ensure that one obtains the correct relic density and an acceptable velocity distribution for $N_1$.

Assuming that such a mechanism is at work, the elastic $N_1$-nucleon cross section is again evaluated using micrOMEGAs. Both the spin independent (SI) and spin dependent (DS) scattering cross sections with nucleons turn out to be constrained by experiments. 

The currently most stringent constraint on SI nucleon scattering comes from the CDMS \cite{Ahmed:2009zw} and Xenon 100 \cite{Aprile:2010um} experiments, and have reached a sensitivity of $4\times 10^{-8}$~pb. The SI cross section is dominated by Higgs boson exchange and thus depends on $m_h$ and on the light quark form factors of the nucleon. 
In the MSSM the strange quark form factor $f_s$ is a major source of uncertainty, since the couplings to down type quarks are enhanced by $\tan\beta$, see e.g. \cite{Giedt:2009mr,Cao:2010ph,GSW} for recent discussions of this issue. 

\begin{figure}
\begin{center}
\includegraphics{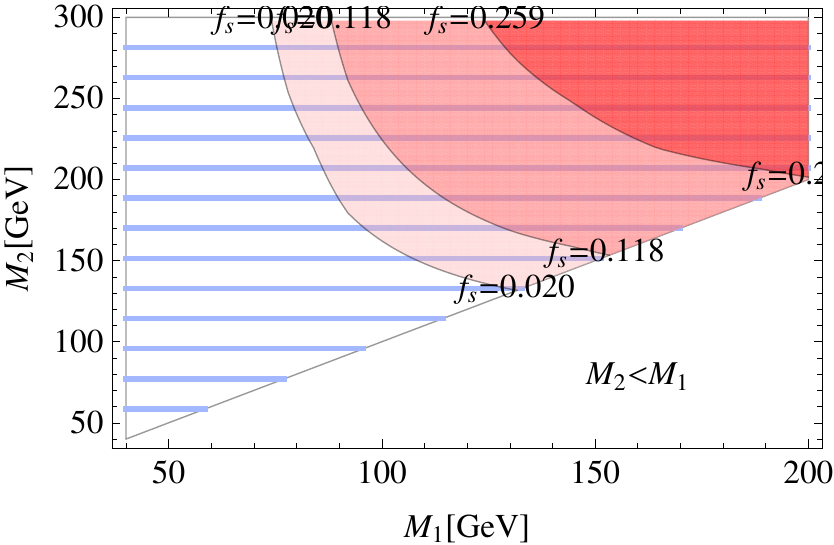}
\caption{Direct detection constraints on $N_1$ dark matter in the $M_1$-$M_2$ plane, for $m_h = 150$~GeV and $M_\ell = M_1 + 200$~GeV, under the assumption that $\Omega_{N_1} = \Omega_{\rm{DM}}$. The red shaded areas are excluded by constraints on the spin dependent nucleon cross section, for the indicated values of $f_s$. The whole parameter space is also excluded by constraints on the spin dependent neutrino-neutron cross section, as indicated by the blue stripes. }
\label{fig:dd}
\end{center}
\end{figure}

For the case considered here, the strange quark form factor is not as dominant, but still affects the exclusion limit, as will be shown later. The parameter dependence of the Higgs mediated scattering cross section is easy to understand. The $N_1$ couplings to the Higgs boson are proportional to $M_1$ and also enhanced for large $M_2$, while the exchange of the Higgs boson in the t-channel gives rise to a $m_h^{-4}$ behavior.  For small $m_h$ this leads to exclusion of $M_1$ values above $150-200$~GeV. In fig. \ref{fig:dd} the excluded region in the $M_1-M_2$ plane is shown for the case of a $150$~GeV Higgs boson. The dependence on the strange quark form factor is illustrated by showing the constraint for three different values of $f_s$.

Constraints on spin-dependent neutrino-nucleon scattering are much weaker when compared naively to the SI constraints, the best limit coming from the Xenon 10 experiment \cite{Angle:2008we}. It constraints the SD neutron cross section to be smaller than $5\times 10^{-3}$~pb. This limit turns out to be very restrictive for Majorana neutrino dark matter, due to its large coupling to the $Z$-boson. A purely Majorana neutrino is excluded in the mass range of $10$~GeV$<M<2$~TeV. 

In the mixed Dirac-Majorana case considered here, the coupling to the $Z$-boson is reduced by $c_\theta^2$, thus the cross section is reduced by $c_\theta^4> \frac{1}{4}$. In the case of maximal mixing, the constraint becomes somewhat weaker, but still excludes Majorana neutrino dark matter for masses roughly between $15$~GeV and $500$~GeV, and thus the whole parameter range in fig. \ref{fig:dd}.

In conclusion, forth generation neutrino dark matter is heavily constrained by dark matter direct detection experiments, even in the mixed Dirac-Majorana scenario, under the assumption that the neutrino relic density matches the observed dark matter abundance. These constraints are relaxed if either the local dark matter density and velocity distribution is different from the one assumed to obtain the limits, or if the Majorana neutrinos are only one of several dark matter species. In particular case where $N_1$ is a thermal relic is not constrained by spin dependent neutron scattering, since the density is at least a factor of 5 smaller. This regime will however be probed by the next generation of experiments~\cite{Freytsis:2010ne}.
Finally $N_1$ annihilation in the sun and in the galactic halo is suppressed by the small thermal relic abundance, and therefore does not impose further constraints on the scenarios considered here~\cite{Fargion:1994me,Belotsky:2008vh}.

\section{Conclusions}
We have analyzed the effects of stable fourth generation Majorana neutrinos on Higgs phenomenology and on cosmology. 

The rate $\sigma_{hWW}=\sigma(pp \to h) \times {\rm Br}(h \to W^+ W^-)$ is an important test of the standard model of physics, and will eventually be measured at the LHC, provided that the Higgs boson is there. A deviation of the SM prediction is a clear indication that new physics is changing either the Higgs production cross section or its branching fractions, or both. We have shown that in the presence of stable Majorana neutrinos, this rate $\sigma_{hWW}$ can be reduced significantly, despite the strong enhancement of the Higgs production cross section due to the fourth generation quarks. 

The semi-invisible decays of the Higgs boson into $N_1N_2$ and $N_2N_2$ pairs provide unique signals at the LHC, and might play an important role in understanding deviations in $\sigma_{hWW}$. When the mass splittings between the neutrino states are small, these signals are very challenging, since the produced leptons are rather soft. 

A stable $N_1$ is not excluded by current dark matter direct detection experiments, provided that $M_2 - M_1$ is large enough to prevent large inelastic scattering rates. In this case $N_1$ will have a nonzero thermal relic density that can reach up to 20\% of the observed dar matter density, but is below 10\% for most of the parameter space. The possibility that all dark matter is of $N_1$ type, e.g. invoking a non-thermal production mechanism, is excluded by constraints on the spin dependent dark matter-neutron cross section. Future direct detection experiments will strengthen these bounds and constrain the fraction of dark matter that can be of $N_1$ type to be less than 10\%. 

A stable fourth generation has an abundance of interesting phenomenological consequences. Besides the effects on Higgs phenomenology presented here, there can be signals of stable charged particles \cite{Murayama:2010xb}, quark-antiquark bound states \cite{Barger:1987xg,Hung:2009hy,Ishiwata:2011ny} and lepton flavor violation \cite{Carpenter:2010bs} at the LHC. In the context of the present discussion, it would be good to investigate further the prospects of the LHC to detect the semi-invisible decay modes of the Higgs boson. 

\section*{Acknowledgements}

We thank H.-S.~Lee, A.~Soni and R.~Vega-Morales for useful discussions, R.~Boughezal for valuable comments on the manuscript, and S.~Gori and C.~Wagner for collaboration on related topics. 
W.-Y.~K. thanks BNL for hospitality.
This work was supported in part  by the U.S. Department of Energy, Division of High Energy Physics, under Contract DE-AC02- 06CH11357 and DE-FG02-84ER40173.

\appendix

\section{Neutrino Couplings in Four Component Formalism}
A 2-component chiral field $N$  is stacked up 
with its Hermitian conjugate $N^\dagger$ to form  
a 4-component Majorana field $N^M=N \oplus N^\dagger$.

Thus  we have 
\begin{align}
 N_1^2 +\hbox{ h.c. } &\longrightarrow \bar N^M_1 \BM{1} N^M_1  \ , \\
N_2^2 +\hbox{ h.c. } &\longrightarrow \bar N^M_2 \BM{1} N^M_2  \ , \\
iN_2N_1 +\hbox{ h.c. } &\longrightarrow 
\bar N^M_2 i\gamma_5  N^M_1  
=\bar N^M_1 i\gamma_5  N^M_2  
\ . \end{align}
As usual $\bar\psi = \psi^\dagger \gamma^0$. Using the above relations it is easy to translate the Higgs boson couplings (\ref{eqn:higgscoupling}) into four component language 
\begin{align}
	{\cal L}  = \frac{1}{2}\frac{m_{4D}}{\sqrt{2}v} h \left (2 c_\theta s_\theta \bar{N}^M_1 N^M_1 +  2 c_\theta s_\theta \bar{N}^M_2N^M_2 +2  (s_\theta^2 - c_\theta^2) \bar{N}^M_1 i \gamma_5 N^M_2 \right),
\end{align}
The four component couplings to gauge bosons are obtained in a similar manner. For this we note that, in the Weyl basis, the $4\times 4$ gamma matrices are defined as 
\begin{align}
	\gamma^\mu & = \begin{pmatrix} 0 & \sigma^\mu \\ \bar\sigma^\mu & 0 \end{pmatrix}, && P_L = \frac{1}{2}(\BM{1} - \gamma_5)  =  \begin{pmatrix} \BM{1} & 0 \\ 0& 0 \end{pmatrix},
\end{align}
such that the lefthanded two component coupling $N_i^\dagger \bar\sigma^\mu N_j$ can be rewritten as 
\begin{align}
	N_i^\dagger \bar\sigma^\mu N_j  & \longrightarrow \bar{N}^M_i \gamma^\mu P_L N^M_j\,.
\end{align}
For $i=j$ only the axial part of the coupling survives. Therefore the couplings to the $Z$~boson translate to
\begin{align} 
{\cal L} \supset \frac{g}{c_w}\frac{1}{2} 
\left[-c_\theta^2 \bar N^M_1\gamma^\mu  \frac{\gamma_5}{2} N^M_1
                  -s_\theta^2 \bar N^M_2 \gamma^\mu\frac{\gamma_5}{2}  N^M_2
  +s_\theta c_\theta 
i\bar N^M_1\gamma^\mu  N^M_2 \right] Z_\mu \ .
\end{align}

Further useful relations between the masses $M_{1,2}$ and the quantities $c_\theta$, $s_\theta$ are collected here:
$$ \tan\theta={\sin\theta\over\cos\theta}={m_{4D}\over M_2}
={M_1\over m_{4D}}=\sqrt{M_1\over M_2}\,,  $$
$$ \sin\theta={M_1\over \sqrt{M_1^2+m^2_{4D}}}=\sqrt{M_1\over{M_1+M_2}}
\ ,\quad
   \cos\theta={m_{4D}\over \sqrt{M_1^2+m^2_{4D}}}=\sqrt{M_2\over{M_1+M_2}}
\ ,$$ 
$$ \sin\theta\cos\theta={\sqrt{M_1M_2}\over M_1+M_2 } \ ,\quad
   \cos^2\theta-\sin^2\theta={M_2-M_1\over M_1+M_2} \ .$$
Finally the triangle function $\lambda$ that appears in the two body phase space integrals is defined as
\begin{align}
	\lambda\left(\frac{M_1}{m_h},\frac{M_2}{m_h}\right) =  \left( 1-{M_1^2\over m_h^2}- {M_2^2\over m_h^2}\right)^2
-    4{M_1^2\over m_h^2}{M_2^2\over m_h^2}\,.
\end{align}

\end{document}